\documentclass[11pt]{article}
\usepackage{ams}
\title{Mass, Impetus and Force by Symplectic Realizations of the Static Group}
\author{Joachim Nzotungicimpaye\footnote{On absence from the University of Burundi}\\Kigali Institute of Education, Department of Mathematics\\ P.O.Box 5039,Kigali-Rwanda\\e-mail kimpaye @kie.ac.rw}

\begin{document}
\maketitle
\begin{abstract}
We show by symplectic realizations of the one dimensional Static
group $G$ that the maximal $G$-elementary system is a a massive
particle under an invariant force $f$ participating in the linear
momentum and an invariant impetus $I$ participating in the change
of position. Moreover the system is characterized by four physical
quantities : a mass $m$ ,a boost $u=\frac{I}{m}$, an acceleration
$a=\frac{f}{m}$ and an internal energy $U$. The hamiltonian of the
system is a sum of the potential energy $V=-fq$ and a contribution
$pu$ of the impulse.
\end{abstract}
\section{Introduction}
Following \cite{sanjuan}, there are eleven physically plane
geometries governed by the eleven Lie algebras. Following
\cite{lev1} and \cite{sanjuan} these algebras are generated by $K$
for the boosts, $P$ for space translations and $E$ for time
translations whose physical dimensions are respectively that of an
inverse of a velocity , an inverse of a length and an inverse of a
duration . We distinguish within these Lie algebras
\begin{itemize}
\item two simple Lie algebras which are the de Sitter Lie algebras
${\cal}{dS}_{\pm}$ and defined by the Lie brackets
\begin{eqnarray}
[K,P]=\frac{E}{c^2}~~,~~[K,E]=P~~,~~[P,E]=\pm \omega^2K
\end{eqnarray}
where $c$ is a constant velocity while $\omega$ is a constant
frequency.
 \item five solvable Lie algebras which are
 \subitem the two Newton-Hooke
Lie algebras ${\cal}{N}_{\pm}$ defined by the Lie brackets
\begin{eqnarray}
[K,P]=0~~,~~[K,E]=P~~,~~[P,E]=\pm \omega^2K
\end{eqnarray}
\subitem the Poincar\'{e} Lie algebra ${\cal}{P}$ defined by the
Lie brackets
\begin{eqnarray}
[K,P]=\frac{E}{c^2}~~,~~[K,E]=P~~,~~[P,E]=0
\end{eqnarray}
and \subitem  the two Para-Poincar\'{e} Lie algebras
${\cal}{P}_{\pm}$ defined by the Lie brackets
\begin{eqnarray}
[K,P]=\frac{E}{c^2}~~,~~[K,E]=0~~,~~[P,E]=\pm \omega^2K
\end{eqnarray}
\item three nilpotent Lie algebras which are \subitem the Galilei
Lie algebra $\cal{G}$ defined by the Lie brackets
\begin{eqnarray}
[K,P]=0~~,~~[K,E]=P~~,~~[P,E]=0
\end{eqnarray}
\subitem the Carroll Lie algebra $\cal{C}$ defined by the Lie
brackets
\begin{eqnarray}
[K,P]=\frac{E}{c^2}~~,~~[K,E]=0~~,~~[P,E]=0
\end{eqnarray}
and \subitem the Para-Galilei Lie algebras $\cal{G}^{\prime}$
defined by the Lie brackets
\begin{eqnarray}
[K,P]=0~~,~~[K,E]=0~~,~~[P,E]=\omega^2K
\end{eqnarray}
\item An abelian Lie algebra which is the static Lie algebra
${\cal}{St}$ for which all the above Lie brackets vanish .
\end{itemize}
Note that a general element of a one parameter Lie group generated
by $X$ take the form $exp(xX)$.This means that $xX$ must be
dimensionless and then that the physical dimension of the
parameter $x$ must be be the inverse of that of $X$. For that
reason the parameters $v$, $x$ and $t$ associated respectively to
$K$, $P$ and $E$ will have the dimensions of a velocity, a length
and a duration. Their duals will be a static momentum $k$, a
linear momentum $p$ and an energy $e$. The relations between this
three physical quantities depend on the structure of the Lie
algebras and can be determined by the study of the strongly
hamiltonian realizations of the corresponding Lie groups
(\cite{nzo88} and references therein). This paper study the Static
group case which has never been studied. In two other papers we
study the nilpotent and the solvable cases. One will study the
Carroll group and the Para-Galilei group and will compare the
results with those obtained for the Galilei group that we revisit
for completeness. The other will study the Para-Poincar\'{e}
groups which also have not been studied and compare the results
with those obtained for the Newton-Hooke (\cite{gadella} and
references therein)and the Poincar\'{e} groups \cite{nzo2003} that
also we revisit for completeness.
\section{The Static group and its central extension}
As said in the previous section, the one spatial dimensional
Static Lie algebra $\cal{G}$ is the abelian one generated by $K$
for the boosts , $P$ for space translations and $E$ for time
translations.  The general element of the connected Static group
$G$ can be written a $exp(vK+xP+tE)$ where the parameters $v$ ,
$x$ and $t$ are respectively the velocity parameter, the space
translations parameter and the time translations parameter. We can
represent the element of $G$ by the triplet $g=(v,x,t)$ and the
multiplication law for the static group is then
\begin{eqnarray}\label{law1}
(v,x,t)(v^{\prime},x^{\prime},t^{\prime})=(v+v^{\prime},x+x^{\prime},t+t^{\prime})
\end{eqnarray}
which means that $G=(\mathbf{R}^3,+)$.\\We can verify that the
central extension $\cal{\widehat{G}}$ of the Static Lie algebra is
generated by $K$, $P$, $E$ , $M$, $F$ and $Y$ such that
\begin{eqnarray}\label{bracket}
[K,P]=M~,~[K,E]=Y~,~[P,E]=F
\end{eqnarray}
 This means that $M$ and $F$ have
$L^{-2}T$ and $L^{-1}T^{-1}$ as physical dimension while the
dimension of $Y$ is that of $P$. Let us write the general element
of the connected central extension $\hat{G}$ of the group $G$ as
$exp(\xi M+\zeta F+yY)exp(vK+xP+tE)$ and let us represent the
general element of $\hat{G}$  by $\hat{g}=(\alpha,g)$ where
$\alpha =(\xi,\zeta,y)$. Use of the Baker-Campbell-Hausdorff
\cite{barut} give rise to that the multiplication law for
$\hat{G}$
\begin{eqnarray}\label{law2}
(\alpha,g)(\alpha^{\prime},g^{\prime})=(\alpha+\alpha^{\prime}+c_1(g,g^{\prime}),gg^{\prime})
\end{eqnarray}
where the cocycle $c_1$ is such that
\begin{eqnarray}
2c_1(g,g^{\prime})=(vx^{\prime}-v^{\prime}x,xt^{\prime}-t^{\prime}x,vt^{\prime}-v^{\prime}t
)
\end{eqnarray}
Definition of $b:G\rightarrow \mathbf{R}^3$ by
\begin{eqnarray}
b(g)=\frac{1}{2}(vx,xt,vt)
\end{eqnarray}
permit us to verify that the cocycle $c_2$ given by
\begin{eqnarray}
2c_2(g,g^{\prime})=
(vx^{\prime}+v^{\prime}x,xt^{\prime}+t^{\prime}x,vt^{\prime}+v^{\prime}t)
\end{eqnarray}
is trivial \cite{nzo88} and \cite{nzo94}. The cocycle $c_1$ is
then equivalent to $c=c_1+c_2$ given by
\begin{eqnarray}
c(g,g^{\prime})=(vx^{\prime},xt^{\prime},vt^{\prime})
\end{eqnarray}
and the multiplication law (\ref{law2}) is equivalent to
\begin{eqnarray}\label{law3}
(\alpha,g)(\alpha^{\prime},g^{\prime})=(\alpha+\alpha^{\prime}+c(g,g^{\prime}),gg^{\prime})
\end{eqnarray}
which is explicitly \\
\\
$(\xi,\zeta,\eta,v,x,t)(\xi^{\prime},\zeta^{\prime},\eta^{\prime},v^{\prime},x^{\prime},t^{\prime})$
\begin{eqnarray}\label{law4}
=(\xi+\xi^{\prime}+vx^{\prime},\zeta+\zeta^{\prime}+xt^{\prime},y+y^{\prime}+vt^{\prime},v+v^{\prime},x+x^{\prime},t+t^{\prime})~~~~~~~~~~~~~
\end{eqnarray}
\section{$G-$elementary Systems}
We verify from (\ref{law4}) that the adjoint action of $G$ on the
central extension $\cal{\widehat{G}}$ of $\cal{G}$ is\\
\\
$Ad_{(v,x,t)}(\delta \xi, \delta \zeta, \delta y,\delta v, \delta
x , \delta t)$
\begin{eqnarray}\label{adjoint}
=(\delta \xi+v\delta x-x\delta v, \delta \zeta +x\delta t-t\delta
x, \delta y+v\delta t-t\delta v,\delta v, \delta x, \delta
t)~~~~~~~~~
\end{eqnarray}
If the duality is defined by\\
\\
$<(m,f,I,k,p,e),(\delta \xi, \delta \zeta, \delta y,\delta v,
\delta x , \delta t)>$
\begin{eqnarray}\label{duality}
=m\delta \xi+f \delta \zeta+I \delta y+k\delta v+p \delta x +e
\delta t~~~~~~~~~~~~~~~~~~~~~~~~~~~~~~~~~~
\end{eqnarray}
and if the right hand side of (\ref{duality}) has action as
physical dimension, then $m$,$f$, $I$,$k$,$p$ and $e$ have
respectively the physical dimensions of a mass, a force, an
impetus, a static momentum (mass times position) ,a linear
momentum
and an energy.\\
We then verify that the coadjoint action of $G$ on the dual
$\cal{\widehat{G}^*}$ of $\cal{\widehat{G}}$

is given by\\
\\
$Ad^*_{(v,x,t)}(m,f,I,k,p,e)$
\begin{eqnarray}
=(m,f,I,k+mx+It,p-mv+ft,e-fx-Iv)~~~~~~~~~~~~~~~~~~~~~~~~
\end{eqnarray}
Note that the impetus $I$ produces the energy $Iv$ and the static
momentum $It$ .It does not produces linear momentum. \\ We verify
from (\ref{bracket}) that the Kirillov form is
\begin{eqnarray}
K_{ij}(m,I,f)=\left (
\begin{array}{ccc}
0&m&I\\-m&0&f\\-I&-f&0
\end{array}
 \right)
\end{eqnarray}
Each orbit being an elementary system, we distinguish eight kind
of elementary systems. Four massive ones corresponding to $m\neq
0$ and four massless ones corresponding to $m=0$. Among the
massive ones as well among the massless ones there two accelerated
systems and two free systems. Moreover we distinguish among the
accelerated as well among the free ones , two boosted systems and
two static systems.
\section{Physics of the orbits}
In this section we study the physics of the eight elementary
systems associated to the Static group.
\subsection{ Massive Systems }
These systems correspond to the case $m\neq 0$. Among them we
distinguish those which are accelerated ($f\neq 0$) from the free
ones ($f=0$).
\subsubsection{Accelerated Boosted Massive Systems "ABS"}
This system corresponds to the case $I\neq 0$. It is characterized
by four invariants, three of them $m$,$f$ and $I$ being trivial,
the fourth one $U$ being
\begin{eqnarray}
U=e-pu+fq\equiv e-pu+ka
\end{eqnarray}
where
\begin{eqnarray}\label{cc}
u=\frac{I}{m}~~,q=\frac{k}{m}~~,a=\frac{f}{m}
\end{eqnarray}
Note that $u$ is a boost while $a$ is an acceleration.\\
One also verify that the coadjoint orbit is a symplectic manifold
endowed with the symplectic form
\begin{eqnarray}\label{sympform1}
\sigma=dp \wedge dq
\end{eqnarray}
 The action of the static group on
$C^{\infty}({\cal{O}}_{(m,~f,~I,~U)},\mathbf{R})$ is given by
\begin{eqnarray}\label{symplrealization1}
(D_{(v,~x,~t)}\psi)(p,q)=\psi(p+mv-ft, q-\frac{I}{m}t-x )
\end{eqnarray}
We see that the force $f$ participate in the linear momentum while
the impetus $I$ participate in the change of position. We then
verify that the Static Lie algebra is realized by the hamiltonian
vector fields
\begin{eqnarray}
D(K)=m\frac{\partial}{\partial p}~~,~~D(P)=-\frac{\partial
}{\partial q}~~,~~D(E)=-f\frac{\partial}{\partial
p}-u\frac{\partial }{q}
\end{eqnarray}
The components of the corresponding momentum are then a static
momentum
\begin{eqnarray}
\mu(K)=mq,
\end{eqnarray}
a linear momentum
\begin{eqnarray}
\mu(P)=p,
\end{eqnarray}
and an energy
\begin{eqnarray}\label{energy}
\mu(E)=pu-fq
\end{eqnarray}
We verify also verify from (\ref{symplrealization1}) that the
motion equations are
\begin{eqnarray}
f=\frac{dp}{dt}~,~~I=m\frac{dq}{dt}
\end{eqnarray}
which are the usual definition of force and impetus by Newton
laws. We also see that the hamiltonian function is up an additive
constant
\begin{eqnarray}\label{hamiltonian}
H(p,q)=pu-fq
\end{eqnarray}
The relation (\ref{hamiltonian}) shows that the hamiltonian is a
sum of a kinetic contribution $pu$ of the impetus and a potential
energy $-fq$.\\
{\it The orbit describes then a constantly boosted massive
particle under a constant force $f$}.
\subsubsection{Accelerated massive static system "ASS"}
The difference with the previous case is that $I=0$ and then that
the internal energy is
\begin{eqnarray}
U=e-fq
\end{eqnarray}
We can then denote the orbit by ${\cal{O}}_{(m,f,U)}$. It is still
endowed with (\ref{sympform1}) and the action of the Static group
on $C^{\infty}({\cal{O}}_{(m,f,U)},\mathbf{R})$ is
\begin{eqnarray}\label{symplrealization2}
(D_{(v,~x,~t)}\psi)(p,q)=\psi(p+mv-ft, q-x )
\end{eqnarray}
The motion equations are now
\begin{eqnarray}\label{eqmotion2}
f=\frac{dp}{dt}~,~~\frac{dq}{dt}=0
\end{eqnarray}
while the hamiltonian is, up an additive constant, the potential
energy
\begin{eqnarray}
H(p,q)=-fq
\end{eqnarray}
From (\ref{eqmotion2}) we can say that {\it the orbit is a static
massive particle under a constant force $f$}.
\subsubsection{Boosted Free Massive System "BFS"} This case corresponds to
$f=0$ and $I\neq  0$. The difference with the accelerated massive
system with impetus is that $f=0$. It is then characterized by the
mass $m$, the impetus $I$ and the internal energy
 \begin{eqnarray}
 U=e-pu
\end{eqnarray}
We can denote the orbit by ${\cal{O}}_{(m,I,U)}$ .It is endowed
with (\ref{sympform1}) and the action of the static group on it is
such that
\begin{eqnarray}\label{symplrealization3}
(D_{(v,~x,~t)}\psi)(p,q)=\psi(p+mv, q-ut-x )
\end{eqnarray}
The motion equations are
\begin{eqnarray}\label{motion3}
\frac{dp}{dt}=0~~,~~I=m\frac{dq}{dt}
\end{eqnarray}
while the Hamiltonian is up an additive constant
\begin{eqnarray}\label{ham3}
H(p,q)=pu
\end{eqnarray}
which is a kinetic energy.\\{\it The orbit describes a constantly
boosted free massive particle }.
\subsubsection{Free Massive Static System "FSS"}
 In this case the energy $e$ dual to time translations is a trivial
 invariant and the orbit can be denoted ${\cal{O}}_{(m,e)}$. It is also endowed with (\ref{sympform1})
 and the action of the Static group on it is such that
\begin{eqnarray}\label{symplrealization4}
(D_{(v,~x,~t)}\psi)(p,q)=\psi(p+mv, q-x )
\end{eqnarray}
The motion equations are
\begin{eqnarray}\label{eqmotion4}
\frac{dp}{dt}=0~~,~~\frac{dq}{dt}=0
\end{eqnarray}
while the hamiltonian is the invariant one $H=e$. The equation
(\ref{eqmotion4}) tell us that {\it the orbit is a free static
particle}.
\subsection{Massless Systems}
These systems correspond to the cases where $m=0$.As for the
massive systems we distinguish among them those which are
accelerated ($f\neq 0$) from the non-accelerated ($f=0$).
\subsubsection{Boosted Massless system under a force
"BSF"} In this case the system is characterized by the force $f$,
the impetus $I$ and
\begin{eqnarray}
k_0=k-\frac{p}{\omega}
\end{eqnarray}
where $\omega=\frac{f}{I}$ is an invariant frequency. We can
denote the corresponding orbit by ${\cal{O}}_{(f,I,k_0)}$. It is
endowed with the symplectic (\ref{sympform1}) form with
\begin{eqnarray}\label{position1}
q=-\frac{e}{f}
\end{eqnarray}
The action of the Static group on
$C^{\infty}({\cal{O}}_{(f,I,k_0)},\mathbf{R})$ is
\begin{eqnarray}\label{symplrealization5}
(D_{(v,~x,~t)}\psi)(p,q)=\psi(p-ft,q-\frac{v}{\omega}-x)
\end{eqnarray}
while the motion equations are
\begin{eqnarray}\label{motion5}
\frac{dq}{dt}=0~~,~~f=\frac{dp}{dt}
\end{eqnarray}
We can verify that the hamiltonian is the potential energy
\begin{eqnarray}\label{potentialenergy}
H=-fq
\end{eqnarray}
{\it The orbit describes a boosted massless system under a
constant force $f$.}
\subsubsection{Massless static system under a force "SSF"}
In this case the system is characterize by a invariant force $f$
and $k$. It is endowed with (\ref{sympform1}) and the action of
the Static group on $C^{\infty}({\cal{O}}_{(f,k)},\mathbf{R})$ is
\begin{eqnarray}\label{symplrealization6}
(D_{(v,~x,~t)}\psi)(p,q)=\psi(p-ft,q-x)
\end{eqnarray}
The motion equations are still (\ref{motion5}) while the
hamiltonian is also (\ref{potentialenergy}).\\{\it The orbit
describes a massless static system under a constant force $f$.}
\subsubsection{Boosted Free Massless System "BFS"}
In this case the linear momentum $p$ is a trivial invariant and
the system is now characterized by the impetus $I$ and $p$. We
denote the corresponding orbit by ${\cal{O}}_{(I,p)}$. It is
endowed with the symplectic form
\begin{eqnarray}
\sigma=de \wedge d\tau
\end{eqnarray}
 where
\begin{eqnarray}
\tau=\frac{k}{I}
\end{eqnarray}
The action of the Static group on
$C^{\infty}({\cal{O}}_{(I,p)},\mathbf{R})$ is
\begin{eqnarray}\label{symplrealization6}
(D_{(v,~x,~t)}\psi)(e,\tau)=\psi(e+Iv,\tau-t)
\end{eqnarray}
The motion equations are then
\begin{eqnarray}
\frac{de}{dt}=0~~,~~\frac{d\tau}{dt}=1
\end{eqnarray}
hamiltonian is $H=e$ as in the case of a {\it free boosted
massless particle}.
\subsubsection{Free Massless static system "FSS"}
This is a trivial ($0$-dimensional) system consisting of a point
$(k,p,e)$ which is {\it a free static massless particle} on which
the static group act trivially.
\section{Conclusion}
Let us summarize the results in one table for massive systems and
in another for massless systems.
\begin{itemize}
 \item Massive systems
 \begin{eqnarray*}
 \begin{tabular}{|l|l|l|l|}
\hline system&$(D_{(v,x,t)}\psi)(p,q)$&motion
equations&hamiltonian\\
 \hline
 ABS&$\psi(p+mv-ft,q-ut-x)$&$f=\frac{dp}{dq}~,~I=m\frac{dq}{dt}$&$H=pu-fq$\\
 \hline
 ASS&$\psi(p+mv-ft,q-x)$&$f=\frac{dp}{dq}~,~\frac{dq}{dt}=0$&$H=-fq$\\
 \hline
 BFS&$\psi(p+mv,q-ut-x)$&$\frac{dp}{dq}=0~,~I=m\frac{dq}{dt}$&$H=pu$\\
 \hline
 FSS&$\psi(p+mv,q-x)$&$\frac{dp}{dq}=0~,~\frac{dq}{dt}=0$&$H=e$\\
 \hline
 \end{tabular}
 \end{eqnarray*}
 \item Massless systems
\begin{eqnarray*}
 \begin{tabular}{|l|l|l|l|}
\hline system&$(D_{(v,x,t)}\psi)(p,q)$&motion
equations&hamiltonian\\
 \hline
 BSF&$\psi(p-ft,q-\frac{v}{\omega}-x)$&$f=\frac{dp}{dq}~,~\frac{dq}{dt}=0$&$H=-fq$\\
 \hline
SSF&$\psi(p-ft,q-x)$&$f=\frac{dp}{dq}~,~\frac{dq}{dt}=0$&$H=-fq$\\
 \hline
 BFS&$\psi(e+Iv,\tau-t)$&$\frac{de}{dt}=0~,~\frac{d\tau}{dt}=0$&$H=e$\\
 \hline
 \end{tabular}
 \end{eqnarray*}
 \end{itemize}
 We do not include the Free Massless Stactic System which is a fixed
 point.\\The realizations on the orbits ABS, ASS, BFS and BSF are
 faithful while they are unfaithful on the others.

\end{document}